\documentclass[lettersize,journal]{IEEEtran}

\usepackage[caption=false,font=normalsize,labelfont=sf,textfont=sf]{subfig}
\usepackage{verbatim}
\usepackage{cite}
\usepackage{tikz}
\usepackage{amsmath,amssymb,amsfonts}
\usepackage{algorithmic}
\usepackage{algorithm}
\usepackage{array}
\usepackage{graphicx}
\usepackage{caption}
\usepackage{textcomp}
\usepackage{xcolor}
\usepackage[most]{tcolorbox}
\usepackage{makecell}
\usepackage{enumitem}
\usepackage{multirow}
\usepackage{booktabs}
\usepackage{diagbox}
\usepackage{colortbl}
\usepackage{float}
\usepackage{stfloats}
\usepackage{booktabs}
\newcommand{\needcite}[1]{{\color{red} [needcite]}}

\usepackage{breakurl}           
\usepackage{xurl}                
\definecolor{highlight}{RGB}{200,230,255}
\definecolor{worst}{RGB}{255,200,200}
\newtcolorbox{findingsbox}{
    colback=gray!10,     
    colframe=gray!50,    
    arc=5pt,             
    boxrule=0.8pt,       
    left=5pt, right=5pt, 
    top=5pt, bottom=5pt, 
    boxsep=0pt,          
    breakable            
}

\usepackage{listings}
\definecolor{codebackground}{rgb}{0.95, 0.95, 0.95}
\usepackage{hyperref}
\hypersetup{
  colorlinks=false,  
  pdfborder={0 0 0}
}

\begin{document}

\title{When Code Crosses Borders: A Security-Centric Study of LLM-based Code Translation
}

\author{Hailong Chang, Guozhu Meng, Shuhui Xiao, Kai Chen, Yilin Li, Kun Sun
\thanks{Manuscript created November, 2025. Hailong Chang, Guozhu Meng, Shuhui Xiao, Kai Chen, Yilin Li, Kun Sun are with Institute of Information Engineering, CAS (email: \{changhailong, mengguozhu, xiaoshuhui, chenkai, sunkun2023, liyilin2023\}@iie.ac.cn).}
}

\maketitle

\begin{abstract}
Code translation is crucial for cross-language codebase migration, and large language models (LLMs) have emerged as a promising technique to automate this process. However, the security implications of using LLMs for code translation remain largely unexplored, as existing evaluations primarily focus on syntactic and functional correctness. To bridge this gap, we conduct a security-centric empirical study to investigate the risks of vulnerabilities being introduced or preserved during LLM-based translation. Our study involves a rigorous evaluation of five state-of-the-art LLMs on a curated dataset of 720 security-related code samples across four programming languages (Java, PHP, C, C++) and nine Common Weakness Enumeration (CWE) categories. The results reveal significant security degradation, with 28.6\% to 45\% of translations introducing new vulnerabilities. Web-related flaws, particularly in input validation, proved most challenging for LLMs. Furthermore, we identify and categorize the root causes of these vulnerable translations into a taxonomy of five major error types. Based on our findings, we develop and evaluate a Retrieval-Augmented Generation (RAG)-based mitigation strategy, which successfully reduces the vulnerability introduction rate by 32.8\%. Our study provides the first large-scale evidence of serious security risks in LLM-based code translation and demonstrates the potential of knowledge-enhanced prompting to mitigate them.
\end{abstract}

\section{Introduction}
With the continuous evolution of software systems, the demand for codebase migration has become increasingly critical, such as upgrading enterprise applications~\cite{kalia2021mono2micro, echeverria2015legacy, krishna2021transforming, perez2021software, haugeland2021migrating}. Code translation is widely regarded as one of the fundamental approaches for solving both security~\cite{LinuxKerneltoRust, LinusTorvalds2020} and efficiency~\cite{phptogoKAIROS, phptogoRisbin, phptogoRoelof} challenges in software development. Considering the substantial human effort required for code translation~\cite{githubupgrading}, automated code translation technique can help efficiently migrate projects between Programming Languages (PLs) to meet diverse needs of upgrading.

Recognizing the importance of code translation, various automated techniques have emerged for reliable migration, including rule-based systems~\cite{cxgo, c2rust, weisz2021perfection} and artificial intelligence-driven approaches~\cite{NEURIPS2018_d759175d, nguyen2014migrating, lachaux2021dobf, roziere2020unsupervised}. With the rapid advancement of large language model (LLM) capabilities, LLM-based approaches\cite{yuan2024transagent, ziftci2025migrating} have demonstrated remarkable potential in both accuracy and readability. Empirical evaluations demonstrate that LLMs achieve 70-80\% accuracy in code translation tasks across diverse language pairs~\cite{tao2024unraveling}. 

However, prior research on code translation has mainly focused on benchmarks that emphasize algorithmic problems, with evaluation metrics centered on syntactic accuracy and functional consistency~\cite{yan2023codetransocean, pan2024lost, xue2025classevaltevaluatinglargelanguage}. While these evaluations provide valuable insights into basic translation capabilities, they overlook the security risks in code translation. Insecure translations often propagate vulnerabilities through weakened security controls, creating latent risks that may remain undetected until exploitation~\cite{LLMmistakecost}. These risks highlight the critical need for systematic security evaluation designed for LLM-based code translation, thereby providing significant practical insights for academic and industrial communities.

Security evaluation of LLM-based code translation faces three key challenges. First, there is a lack of datasets tailored for translation security assessment, as existing benchmarks focus largely on algorithmic code without capturing security-critical contexts. Second, current automated vulnerability detection tools suffer from limited accuracy. Third, manual evaluation demands rare expertise in multiple programming language paradigms and their associated security practices. Together, these challenges impede the establishment of strong safety guarantees for code translation in real-world projects.

To address these challenges and enable a systematic security evaluation, we collected and curated a corpus of security-related, file-level code samples for this empirical study. This corpus comprises samples across four programming languages covering nine prevalent Common Weakness Enumeration~\cite{cwedatabase} (CWE) categories. We automatically aggregated vulnerable code files from authoritative sources like CVE~\cite{cvedatabase} and NVD~\cite{nvddatabase} databases, followed by a 180-person-hour manual annotation process to identify security patches and vulnerabilities, while excluding overly complex or incomplete samples. Unlike existing code translation evaluations, our data collection uniquely focuses on security contexts, enabling multidimensional assessment of LLMs' security preservation capabilities during translation.

Our security evaluation framework combines manual analysis with scalable automated techniques to solve the unique challenges of evaluating LLM-based code translation. The \textbf{human-involved evaluation} employs security researchers to manually inspect translated code, and for \textbf{scalable automated evaluation}, we develop an LLM-as-judge system that contains a multi-stage analytical process. We conduct a comprehensive security assessment of five recent LLMs using our collected code samples, comprising 720 security-related cases across multiple PLs. We evaluate model performance through 6,000 translation instances using three metrics: Functional Correctness Rate (FCR), Vulnerability Introduction/Preservation Rate (VIR/VPR). Three security researchers perform detailed failure case categorization and analysis during a 500-person-hour manual investigation. These annotated failure cases serve dual purposes: (1) providing fundamental insights into LLM security limitations, and (2) forming a knowledge base for our Retrieval-Augmented Generation (RAG) mitigation system.

Based on our analysis, we summarize the following key findings: (1) All LLMs show security degradation in translation (VIR 28.6-45\%), though scaling improves resilience. (2) Web vulnerabilities prove most challenging (45.9\% VIR), dominated by input validation (34.9\%) and API mapping (32.7\%) failures. (3) The C-to-Rust translation scenario shows superior results (18.2\% VIR), highlighting how target language safety mechanisms compensate for LLMs' semantic gaps. (4) RAG mitigation cuts vulnerabilities by 32.8\%, proving LLMs need security knowledge activation.

Our key contributions are:
\begin{itemize}[leftmargin=*]
    \item \textbf{A Comprehensive Security-Centric Evaluation of LLM-based Code Translation.} 
    We conduct the first large-scale empirical study focused on security implications in LLM-based code translation, systematically evaluating five state-of-the-art LLMs across 6,000 translation instances. Our evaluation covers 720 security-related code samples spanning four programming languages and nine high-impact CWE categories.
    
    \item \textbf{A Taxonomy of Translation-Induced Vulnerabilities.} 
    Through 500 person-hours of manual analysis, we identify and categorize the root causes of security flaws introduced during translation into a taxonomy of five major error types with 20 detailed patterns, providing foundational insights into LLMs' security limitations.
    
    \item \textbf{Knowledge-Enhanced Mitigation Strategy.} 
    We develop and evaluate a Retrieval-Augmented Generation (RAG)-based mitigation approach that reduces vulnerability introduction rates by 32.8\%, demonstrating the value of contextual security knowledge in improving translation safety.
    
    \item \textbf{Artifacts.} 
    The source code, collected security-centric code samples, manual analysis results, and implementation of our mitigation framework will be made publicly available~\cite{STEDanonymouslink} to support reproducibility and future research.
\end{itemize}

\section{Background \& Related Work}
\subsection{Code Translation} 
Transpilers or compilers use program analysis for code translating, including C2Rust\cite{c2rust}, CxGo\cite{cxgo}, Sharpen\cite{Sharpen} and Java2CSharp\cite{Java2Csharp}. While these tools can efficiently translate code into another PL, they cannot preserve linguistic features and behaviors. \emph{Rule-based methods} treat code translation as a program synthesis task\cite{weisz2021perfection}. This method relies on a predefined rule base, which specifies how elements in the source code (such as keywords, expressions and statements) should be transformed into corresponding structures in the target language. This approach generally ensures consistency and predictability in the translation results. However, it requires significant time and effort to maintain the rule base, and the readability of the translated code is often poor. \emph{Neural network-based methods} mainly treat code translation as a sequence-to-sequence transformation problem. Chen et al.~\cite{NEURIPS2018_d759175d} design a tree-to-tree neural model for code translation. Nguyen et al.~\cite{nguyen2014migrating} use statistical machine translation. Other work leverages deep learning and unsupervised learning\cite{lachaux2021dobf,roziere2020unsupervised,roziere2021leveraging}. \emph{LLM-based methods} directly use LLMs to solve code translation tasks, such as CodeBERT\cite{feng-etal-2020-codebert}, StarCoder\cite{li2023starcoder}, CodeGen\cite{nijkamp2022codegen}, GPT-4\cite{achiam2023gpt}, Qwen2.5-Coder\cite{hui2024qwen2}, Deepseek\cite{liu2024deepseek}. Some researchers also integrate LLMs to support their translation pipelines\cite{yang2024vert,yang2024exploring,yuan2024transagent, ziftci2025migrating}. Notably, several studies have employed LLMs to design sophisticated code translation systems\cite{nitin2025c2saferrust}. However, these approaches are typically tailored to specific language pairs, and reproducing such frameworks for practical code translation poses significant challenges for end users\cite{li2024translating}. In light of this, our work focuses on evaluating the capability of LLMs to perform this task in a direct, out-of-the-box manner. LLMs have shown remarkable performance in preserving semantics and syntax. However, there is no research concerning the possible vulnerabilities introduced by LLMs during code translation. To address this gap, we systematically assess the content safety implications of using LLMs for code translation
\subsection{Evaluation of LLM-based Code Translation}
LLM-based translation has attracted increasing attention from practitioners, which necessitates a comprehensive evaluation for code translation.
There are an emerging number of datasets constructed for evaluation, such as CodeNet\cite{puri2021codenet} and AVATAR\cite{ahmad2021avatar}. Pan et al.\cite{pan2024lost} evaluate five LLMs' code translation capabilities, systematically categorizing both syntactic and semantic errors in the translations. Eniser et al.\cite{eniser2024towards} conduct an evaluation of five LLMs for source-to-Rust translation, providing comprehensive benchmarking results. Xue et al.\cite{xue2025classevaltevaluatinglargelanguage} construct a class-level code translation benchmark to assess recent LLMs’ performance on class-level code translation. Yan et al.\cite{yan2023codetransocean} construct a large-scale benchmark CodeTransOcean for code translation, and evaluate different translating approaches. Li R et al.~\cite{li2024translating} conducted a comprehensive user study on the translation process from C to Rust, investigating the challenges users encounter when utilizing automated translation tools, including those powered by LLMs. Their work presents a complete workflow for secure translation and highlights the potential risks. While user studies can effectively reveal issues arising in practical application scenarios, they may not fully capture the overall capabilities of LLMs in code translation tasks. \emph{Existing evaluations primarily focus on algorithm tasks, often lacking realistic security contexts in their samples. In contrast, we are the first to conduct a systematic evaluation of LLMs' security capabilities in code translation by constructing a file-level, security-centric dataset.}
\section{Data Preparation}
\begin{figure*}[t]
    \centering
    \includegraphics[width=1.02\textwidth, keepaspectratio]{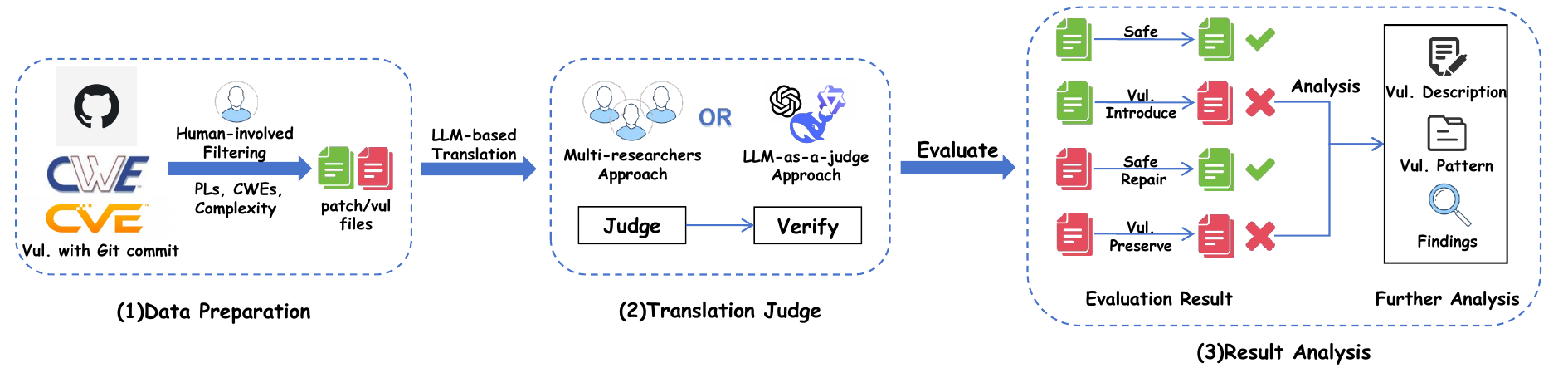}
    \caption{System Overview of Our Approach}
    \label{fig:overview}
\end{figure*}

Fig.~\ref{fig:overview} illustrates the overall methodology of our empirical study, which consists of three main phases. This section details the first phase: the collection and preparation of a security-centric code corpus.

Existing vulnerability datasets are not directly suitable for evaluating security risks in code translation. Function-level datasets~\cite{chen2023diversevul,fan2020ac} often lack complete security context and fine-grained annotations, while file-level datasets~\cite{nikitopoulos2021crossvul,bhandari2021cvefixes} typically collect all available vulnerability-related code changes without systematic filtering. This results in samples that are either excessively long or extremely short, and more critically, they lack the security-specific annotations needed to identify which code portions are directly relevant to vulnerabilities.

Given these limitations and the specific requirements of code translation tasks -- particularly the need for fine-grained security annotations and adequate contextual depth -- we constructed a dedicated code corpus for this study. The objective was to create a foundation for evaluation that adequately represents real-world security scenarios across diverse programming languages and vulnerability types. We describe our target selection criteria and data curation process below.

\subsection{Target Selection}
The data collection for this study aims to comprehensively cover PL diversity, representativeness of CWE types, and hierarchical code complexity. 

Four PLs are selected based on two key criteria: language popularity according to the TIOBE index~\cite{TIOBE} and the distribution of CWE samples across translation pairs~\cite{whitesource}. These criteria lead to the selection of Java, PHP, C and C++ as source languages, with Python, Go, and Rust serving as target languages, thereby forming a diverse set of translation combinations.

In the subsequent analysis, we analyze C and C++ together to ensure sufficient sample size, as both languages exhibit similar safety characteristics, share largely overlapping vulnerability types, and are prone to comparable translation-induced issues. We acknowledge the inherent differences between C and C++, but these factors are accounted for in our evaluation.

As for CWE selection, we refer to the MITRE CWE Top25~\cite{cwetop25}, identifying nine CWE types focusing on high-frequency vulnerability categories such as memory management and input validation, spanning common security scenarios across multiple languages. 

\begin{figure}[t]
    \centering
    \includegraphics[width=0.5\textwidth, height=4.5cm, keepaspectratio]{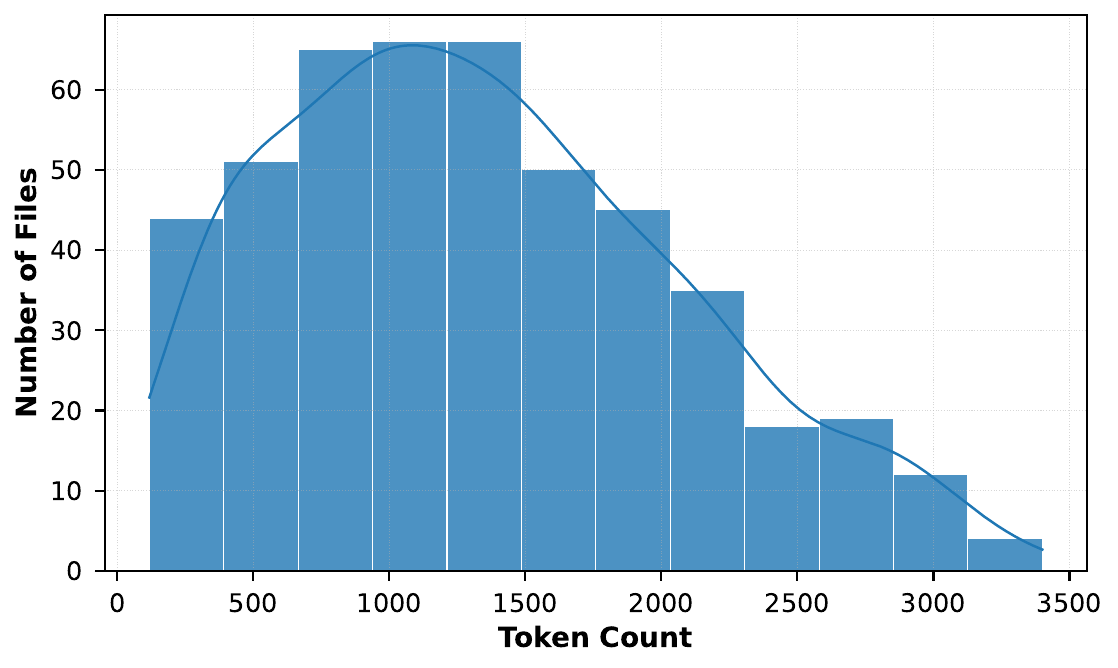}
    \caption{Token Count Distribution of Dataset Files}
    \label{fig:token_dist}
\end{figure}

During preliminary filtering, code complexity is primarily sampled based on token count. As shown in Fig.~\ref{fig:token_dist}, after filtering and refinement, the final code corpus ensures code lengths predominantly ranged between 500 to 1600 tokens while preserving complete security-related code context.

\subsection{Data Collection and Filtering}
\textbf{Data Collection.}
To gather file-level vulnerability data, we leverage the NVD and CVE databases, which maintain extensive records of known vulnerabilities annotated with detailed CWE types and associated code links. Using a custom script, we perform initial filtering based on CWE types and the presence of Git commit links. After further extraction of Git commit data, we screen for programming languages and code complexity, ultimately obtaining 1,726 code samples meeting preliminary criteria.

\begin{table}[t]
\centering
\caption{Dataset Sample Distribution}
\small
\label{tab:dataset}
\begin{tabular}{llrrr}
\toprule
\textbf{Language} & \textbf{CWE Type} & \textbf{Patched} & \textbf{Vulnerable} & \textbf{Subtotal} \\
\midrule
\multirow{3}{*}{C/C++} & CWE-416 & 60 & 60 & 120 \\
                       & CWE-787 & 30 & 30 & 60 \\
                       & CWE-125 & 30 & 30 & 60 \\
                       \cmidrule{2-5}
                       & Total & 120 & 120 & 240 \\
\midrule
\multirow{6}{*}{PHP} & CWE-20 & 31 & 10 & 41 \\
                        & CWE-22 & 20 & 10 & 30 \\
                        & CWE-79 & 40 & 10 & 50 \\
                        & CWE-89 & 30 & 10 & 40 \\
                        & CWE-94 & 30 & 10 & 40 \\
                        & CWE-200 & 40 & 10 & 50 \\
                      \cmidrule{2-5}
                      & Total & 191 & 60 & 251 \\
\midrule
\multirow{6}{*}{Java} & CWE-20 & 29 & 10 & 39 \\
                        & CWE-22 & 40 & 10 & 50 \\
                        & CWE-79 & 20 & 10 & 30 \\
                        & CWE-89 & 30 & 10 & 40 \\
                        & CWE-94 & 30 & 10 & 40 \\
                        & CWE-200 & 20 & 10 & 30 \\
                        \cmidrule{2-5}
                        & Total & 169 & 60 & 229 \\
\midrule
\multicolumn{2}{l}{\textbf{Total}} & \textbf{480} & \textbf{240} & \textbf{720} \\
\bottomrule
\end{tabular}
\end{table}

\textbf{Data Filtering.}
The requirement for semantic integrity in patch-related contexts for code translation tasks exposes inherent limitations in automated filtering methods for preserving critical vulnerability fix contexts. 

Given the typically high complexity distribution of real-world vulnerable code, we adopt a human-driven approach for dataset construction. Security researchers first manually screen real-world cases to ensure samples retain semantic integrity while broadly covering typical fixes for target CWE types. To address complexity hierarchy needs in training data, security experts construct simplified cases based on real-world examples, systematically controlling code complexity while maintaining authentic patch-related contexts. All samples undergo rigorous double-blind review, with two independent researchers evaluating semantic integrity and security fix correctness — only samples achieving consensus are retained.

The final code corpus used in our study comprises 720 code samples of security scenarios, with 88.9\% sourced from real-world repositories and 11.1\% constructed by security researchers to ensure comprehensive coverage of vulnerability patterns. The manual curation process, including data filtering and quality validation, requires 180 person-hours of human effort from our team. We ensure both the semantic integrity of patches and the balance representation across different complexity levels and CWE categories. The specific distribution of the code samples is illustrated in the Table~\ref{tab:dataset}.
\section{Experimental Setup}
This section details the second and third phases of our empirical study methodology shown in Fig.~\ref{fig:overview}: the experimental setup for LLM-based code translation and our hybrid evaluation framework. We design a multidimensional approach to examine how translation impacts code security.

\subsection{LLM-based Code Translation Setup}

\begin{table}[htbp]
  \centering
  \footnotesize
  \caption{Details of the studied LLMs}
  \label{tab:llms}
  \setlength{\tabcolsep}{2pt}
  \begin{tabular}{ccccc}
    \toprule
    Model & Release Date & Size & Arch & Open-source \\
    \midrule
    GPT4omini\cite{gpt4o} & 2024-07 & $\sim$80B & MoE & No \\
    GPT4o\cite{gpt4o} & 2024-05 & $\sim$200B & Dense & No \\
    DeepSeekV3-0324~\cite{liu2024deepseek} & 2025-03 & 671B & MoE & Yes \\
    Qwen2.5MaX-0125~\cite{qwen2.5max} & 2025-01 & $\sim670$B & MoE & No \\
    Qwen2.5Coder~\cite{hui2024qwen2} & 2024-11 & 32B & Code & Yes \\
    \bottomrule
  \end{tabular}
\end{table}

\textbf{LLMs Selection.} 
Table~\ref{tab:llms} shows the selected LLMs. These LLMs are selected based on three key criteria: \emph{state-of-the-art performance in code tasks}, \emph{service availability and stability}, and \emph{diversity in model scale and architectures}.
All selected LLMs, which are released since 2024, demonstrate superior performance on standardized code generation benchmarks (e.g., HumanEval and MBPP)~\cite{liu2024deepseek, xia2024top}. 
Due to experimental constraints, we limit our selection to models with stable API services, ensuring reproducibility. 
Our selection includes both general LLMs and specialized code models. The parameter scale variation is deliberately considered, exemplified by the contrast between lightweight GPT-4o mini and its full-scale counterpart.

\textbf{Program Languages Selection.} 
We establish Java, PHP, and C/C++ as source languages, with Python, Go, and Rust as targets. The translation experiments focus on five strategically selected pairs: Java-to-Python, Java-to-Go, PHP-to-Python, PHP-to-Go, and C/C++-to-Rust. The Java/PHP to Python/Go combinations represent prevalent language migration scenarios in web application development\cite{javatogoVivek, javatopyYMC,phptogoRisbin}, with substantial coverage of CWE vulnerabilities\cite{whitesource}. The C/C++-to-Rust pair embodies a security paradigm shift in systems programming\cite{LinuxKerneltoRust}, featuring abundant memory-related CWE samples. These six language pairs collectively provide comprehensive coverage of security-centric translation scenarios.

\textbf{Translation Setup.}
Building upon our collected security-centric code corpus of 720 samples, we design 1,200 translation tasks for each subject LLM using a baseline prompt template without any enhancements. Each prompt contains: (1) natural language instructions for the translation task; (2) source language; (3) target language and (4) the complete source code to be translated. For file-level code translation in security contexts, we incorporate necessary constraints in the prompts to ensure basic translation quality while avoiding task-specific guidance.

All translation tasks are executed through official model APIs with the following unified configuration:
\begin{itemize}[leftmargin=*]
    \item Temperature: Fixed at 0 to eliminate output randomness
    \item Max tokens: Set to 8,192 to accommodate complete prompt
    \item Top-p: Fixed at 1 to maintain full candidate distribution
\end{itemize}

All other hyperparameters are kept by default, and all evaluations are performed in a zero-shot setting.

\subsection{Code Security Evaluation}
In terms of evaluation methodology, we aim to identify whether vulnerabilities are introduced or preserved during the code translation process. We first test existing vulnerability detection methods and analyze their limitations in this task. To address these challenges, we primarily rely on a rigorous human expert evaluation approach, ensuring the accuracy and reliability of our security assessment.
\begin{table}[htbp]
\centering
\caption{Automated Tools Evaluation (All value in \%)}
\label{tab:tool_performance}
\begin{tabular}{lccc}
\toprule
Tools & Precision & Recall & F1 Score \\
\midrule
CodeQL\cite{codeql_repo} & 77.8 & 29.2 & 42.5 \\
Semgrep\cite{semgrep_repo} & 52.2 & 50.0 & 51.1 \\
Bandit\cite{Bandit} & 50.0 & 66.7 & 57.2 \\
VulnHunter\cite{vulnhuntr} & 73.3 & 44.0 & 55.0 \\
LLM-as-a-judge & 90.5 & 76.0 & 82.6 \\
\bottomrule
\end{tabular}
\end{table}

\textbf{Limitations of Automated Approaches.} We conduct systematic testing on mainstream static analysis tools. We randomly selected 60 Python samples from our translated code corpus to form the test set. Each sample in the test set is manually annotated for the presence of vulnerabilities. All static analysis tools are evaluated using the official rules provided by their developers. As shown in Table~\ref{tab:tool_performance}, these tools achieve less than 60\% F1 score in our task. These findings are consistent with recent studies in the academic community\cite{lenarduzzi2023critical}, reinforcing the conclusion that current static analysis techniques are insufficient for automated evaluation. Although test suites like Juliet\cite{black2018juliet} support automated evaluation in theory, their reliance on executable code and complex multi-file setups, along with the significant effort needed to adapt them to multiple languages, makes them impractical for our case. Additionally, the unpredictable way vulnerabilities arise during translation limits the achievable test coverage.

\textbf{Human-as-a-judge Approach.} Given the limitations of existing automated evaluation methods, we employ a security researcher-driven assessment as the primary evaluation method for our empirical study. Each code translation case undergoes meticulous review by experienced security researchers. The evaluation materials comprise the source code, corresponding CWE types, patch or vulnerability location, the translated code, and associated CVE records.

To ensure consistent and standardized evaluation, we established explicit vulnerability classification criteria through discussions among senior security researchers. All evaluators were trained on these criteria and required to classify identified vulnerabilities according to our predefined taxonomy.

The multi-level review mechanism operates as follows: each translated code sample is independently evaluated by two security researchers. Both researchers perform their assessments without knowledge of each other's conclusions, providing explicit justifications for their decisions. In cases where the two initial evaluators disagree, a third researcher reviews the code and annotations to decide a final result.

Upon completion of all individual assessments, a random sampling validation phase is initiated. Specifically, 10\% of the evaluated cases are randomly selected and re-examined by the review team. This process serves as an additional quality control measure to verify the accuracy and consistency of the initial evaluations, thereby further enhancing the trustworthiness of the outcomes.


\textbf{LLM-as-a-Judge for Reference.} 
To support future research and provide a scalable alternative for preliminary assessment, we also explore an LLM-as-a-judge approach. This automated system employs a multi-stage validation pipeline: multiple LLMs first independently analyze translated code for functional correctness and security vulnerabilities; when discrepancies occur, an arbitration mechanism using DeepSeek-R1~\cite{guo2025deepseek} serves as the final arbiter to resolve conflicts. To enhance assessment accuracy, we incorporate a one-shot prompting strategy that provides contextual demonstrations based on CWE types. As shown in Table~\ref{tab:tool_performance}, this approach achieves an F1 score of 82.6\%. While this is superior to other static analysis tools, it still falls short of the accuracy required for our rigorous empirical study. 

It is important to note that this automated approach \emph{is not used} for the primary evaluations and findings reported in this study. We include it as a supplementary method that may benefit researchers seeking efficient ways to obtain approximate security assessments of LLM-based code translation, particularly when human expert resources are limited. All key results and conclusions presented in this paper are based exclusively on the human expert evaluation described above.
\begin{table*}[ht]
\centering
\caption{Overall Performance of STED Dataset Translation (All values in \%)}
\label{tab:overall_performance}
\small
\setlength{\tabcolsep}{3.5pt}
\renewcommand{\arraystretch}{1.0}
\begin{tabular}{@{}l *{6}{ccc} c @{}}
\toprule
\multirow{2}{*}{\textbf{Models}} & 
\multicolumn{3}{c}{Java→Python} & 
\multicolumn{3}{c}{Java→Go} & 
\multicolumn{3}{c}{PHP→Python} & 
\multicolumn{3}{c}{PHP→Go} & 
\multicolumn{3}{c}{C/C++→Rust} &
\multicolumn{3}{c}{Total} \\
\cmidrule(lr){2-4} \cmidrule(lr){5-7} \cmidrule(lr){8-10} \cmidrule(lr){11-13} \cmidrule(lr){14-16} \cmidrule(l){17-19}
& FCR & VIR & VPR & FCR & VIR & VPR & FCR & VIR & VPR & FCR & VIR & VPR & FCR & VIR & VPR & FCR & VIR & VPR \\
\midrule
GPT4omini & 54.4 & 50.9 & 70.0 & 52.7 & 47.9 & 66.7 & 59.2 & 50.3 & 76.7 & 55.5 & 46.6 & 73.3 & 66.7 & 21.7 & 62.5 & 57.1 & 45.0 & 68.6 \\
GPT4o & 64.5 & 42.0 & 68.3 & 60.9 & 37.9 & 63.3 & \cellcolor{highlight}68.6 & 37.2 & 75.0 & 64.4 & 34.6 & 73.3 & 69.2 & 16.7 & \cellcolor{highlight}48.3 & 65.4 & 34.8 & 62.8 \\
DeepseekV3 & 69.2 & 37.3 & \cellcolor{highlight}63.3 & \cellcolor{highlight}66.9 & 30.8 & \cellcolor{highlight}58.3 & 63.4 & 43.5 & \cellcolor{highlight}71.7 & 63.4 & 35.6 & \cellcolor{highlight}66.7 & 70.8 & 17.5 & 54.2 & 66.3 & 34.2 & \cellcolor{highlight}61.4 \\
Qwen2.5Max & \cellcolor{highlight}69.8 & \cellcolor{highlight}34.9 & 68.3 & 64.5 & \cellcolor{highlight}30.2 & 65.0 & 66.0 & \cellcolor{highlight}30.9 & 73.3 & \cellcolor{highlight}67.5 & \cellcolor{highlight}28.8 & 70.0 & \cellcolor{highlight}74.2 & \cellcolor{highlight}13.3 & 48.3 & \cellcolor{highlight}68.0 & \cellcolor{highlight}28.6 & 62.2 \\
Qwen2.5Coder & 57.4 & 47.9 & 76.7 & 55.6 & 45.6 & 75.0 & 57.6 & 45.5 & 81.7 & 59.2 & 42.4 & 80.0 & 71.7 & 21.7 & 59.2 & 59.5 & 41.9 & 71.9 \\
\addlinespace
\midrule
\rowcolor{gray!10}
Avg & 63.1 & 42.6 & 69.3 & 60.1 & 38.5 & 65.7 & 62.9 & 41.5 & 75.7 & 62.0 & 37.6 & 72.7 & 70.5 & 18.2 & 54.5 & 63.3 & 36.9 & 65.4 \\
\bottomrule
\end{tabular}
\footnotesize
\begin{itemize}[leftmargin=*,noitemsep]
\item[$\ast$] Table cells with \colorbox{highlight}{blue} background denote the best performance in terms of metrics.
\end{itemize}
\end{table*}
\subsection{Evaluation Metrics}
To assess the safety aspects of LLMs in code translation, we refer to evaluation metrics from prior work on LLM-based code translation \cite{yang2024exploring, xue2025classevaltevaluatinglargelanguage} to measure two main aspects, introducing Functional Correctness Rate (FCR) and Vulnerability Introduction Rate (VIR) for security scenarios:

\textbf{FCR} measures the proportion of translated code samples that preserve functional equivalence with their source counterparts. Formally, we define FCR as:
\begin{align}
FCR &=  \frac{1}{N_s}\sum_{i=1}^{N_s} {Cmp}_{\text{func}}(\hat{s_i}, \hat{t_i}) \\
\text{where} \quad {Cmp}_{\text{func}}(s,t) &= 
\begin{cases} 
1 & \text{if } \text{translation is functional correct} \nonumber \\
0 & \text{otherwise} \nonumber
\end{cases}
\end{align}
where $N_s$ denotes the number of translated samples, $\hat{s_i}$ represents the $i$-th source code segment, and $\hat{t_i}$ corresponds to the $i$-th translated code via the target LLM. The comparison function $\text{Cmp}_{\text{func}}(\cdot, \cdot)$ evaluates the functional equivalence during translation.

\textbf{VIR} measures the proportion of samples where security measures are degraded during translation. For source code containing vulnerabilities, we use \textbf{VPR} instead to measure the proportion of translations that preserve the original vulnerabilities. Formally, we define VIR/VPR as:

\begin{align}
VIR &=  \frac{1}{N_p}\sum_{i=1}^{N_p} {Cmp}_{\text{sec}}(\hat{s_i}, \hat{t_i}) \\
\text{where} \quad {Cmp}_{\text{sec}}(s,t) &= 
\begin{cases} 
1 & \text{if } \text{sec}(\hat{s_i}) \neq \text{sec}(\hat{t_i}) \nonumber \\
0 & \text{if } \text{sec}(\hat{s_i}) = \text{sec}(\hat{t_i}) \nonumber
\end{cases}
\end{align}
\begin{align}
VPR &=  \frac{1}{N_v}\sum_{i=1}^{N_v} {Cmp}_{\text{sec}}(\hat{s_i}, \hat{t_i}) \\
\text{where} \quad {Cmp}_{\text{sec}}(s,t) &= 
\begin{cases} 
1 & \text{if } \text{sec}(\hat{s_i}) = \text{sec}(\hat{t_i}) \nonumber \\
0 & \text{if } \text{sec}(\hat{s_i}) \neq \text{sec}(\hat{t_i}) \nonumber
\end{cases}
\end{align}
where $N_p$ denotes the number of patched source samples. $N_v$ denotes the number of vulnerable source samples. The function $\text{sec}(\cdot)$ denotes a binary classification outcome regarding the presence of security vulnerabilities within the code segment. The comparison function $\text{Cmp}_{\text{sec}}(\cdot, \cdot)$ evaluates the security equivalence during translation.
\section{Experimental Results}
Through our experiments, we aim to answer to following Research Questions (RQs).

\textbf{RQ1 (Overall Security Evaluation)}: How accurately do state-of-the-art LLMs preserve security properties in file-level code translation? How does translation accuracy vary across diverse CWE categories? To what extent does code complexity impact vulnerability rates?

\textbf{RQ2 (Vulnerable Translation Patterns)}: What are the different types of underlying causes for translation-introduced vulnerabilities? What is the empirical distribution of these security-specific error types across CWEs?

\textbf{RQ3 (Vulnerable Translation Causes and Risks)}: What root causes in LLM-based code translation lead to the introduction or preservation of vulnerabilities?  How do developers perceive the practical risks and challenges associated with translation-induced vulnerabilities?

\textbf{RQ4 (Mitigation Effectiveness)}: To what extent do the prompt-engineering techniques reduce the risks?

\subsection{RQ1:Overall Security Evaluation}
\textbf{Comparison among LLMs.}
Our evaluation of security-centric code translation performance, as detailed in Table~\ref{tab:overall_performance}, reveals distinct security capabilities across models. Qwen2.5Max, DeepseekV3, and GPT4o collectively achieve superior security-functionality balance (Avg FCR=66.6\%,  VIR=30.5\%, VPR=62.1\%) compared to smaller counterparts. Qwen2.5Max establishes security preservation leadership through best-in-class VIR (28.6\%) and FCR (68.0\%), particularly excelling in memory-critical C/C++→Rust translations (VIR=13.3\%). This performance suggests effective integration of security constraints during translation while maintaining functional correctness. DeepseekV3 demonstrates specialized proficiency in vulnerability prevention, achieving the lowest aggregate VPR (61.4\%) through conservative translation strategies, particularly evident in Java→Go (VPR=58.3\%) and PHP→Python (VPR=71.7\%) tasks. However, this security-first paradigm incurs FCR reductions, revealing inherent correctness-preservation tradeoffs. GPT4o achieves competitive security metrics (VIR=34.8\%, VPR=62.8\%) despite its relatively compact 200B parameter configuration.

In comparisons between code-specialized and general models at comparable parameter scales, Qwen2.5Coder achieves better FCR (59.5\% vs 57.1\%) and lower VIR (41.9\% vs 45.0\%) relative to GPT4omini, but exhibits markedly higher VPR (71.9\% vs 68.6\%). This VPR gap highlights divergent architectural priorities: code-specialized models focus on task fidelity as evidenced by FCR advantages, while general models demonstrate better implicit security awareness through lower VPR.
\begin{findingsbox}
\noindent\textbf{Finding 1:} Larger parameter-scale models generally achieve better security preservation in code translation, with Qwen2.5Max demonstrating superior security preservation (VIR=28.6\%, FCR=68.0\%) and DeepseekV3 achieving the lowest VPR (61.4\%). Code-specialized models improve FCR and VIR but increase VPR, suggesting that domain-specific training does not guarantee better overall security preservation.
\end{findingsbox}
\begin{table*}[ht]
\centering
\caption{Different CWE Performance of STED Dataset Translation (All values in \%)}
\label{tab:cwe_performance}
\small
\setlength{\tabcolsep}{3.5pt}
\renewcommand{\arraystretch}{1.0}
\begin{tabular}{ll *{6}{ccc} c }
\toprule
\multirow{2}{*}{\raisebox{-0.5\height}{\textbf{CWEs}}} & 
\multicolumn{3}{c}{GPT4omini} & 
\multicolumn{3}{c}{GPT4o} & 
\multicolumn{3}{c}{DeepseekV3} & 
\multicolumn{3}{c}{Qwen2.5Max} & 
\multicolumn{3}{c}{Qwen2.5Coder} &
\multicolumn{3}{c}{Total} \\
\cmidrule(lr){2-4} \cmidrule(lr){5-7} \cmidrule(lr){8-10} \cmidrule(lr){11-13} \cmidrule(lr){14-16} \cmidrule(l){17-19}
& FCR & VIR & VPR & FCR & VIR & VPR & FCR & VIR & VPR & FCR & VIR & VPR & FCR & VIR & VPR & FCR & VIR & VPR \\
\midrule
CWE-20 & 63.3 & 41.7 & 80.0 & \cellcolor{highlight}71.7 & 31.7 & \cellcolor{worst}80.0 & \cellcolor{highlight}80.0 & 25.0 & 75.0 & \cellcolor{highlight}76.7 & 21.7 & 75.0 & 65.0 & 38.3 & 90.0 & 71.3 & 31.7 & \cellcolor{worst}81.0 \\
CWE-22 & 55.0 & 48.3 & 70.0 & 63.3 & 40.0 & 65.0 & 63.3 & 38.3 & 75.0 & 68.3 & 28.3 & 75.0 & 60.0 & 43.3 & 80.0 & 62.0 & 39.7 & 73.0 \\
CWE-79 & \cellcolor{worst}48.3 & \cellcolor{worst}56.7 & 65.0 & \cellcolor{worst}56.7 & \cellcolor{worst}43.3 & 60.0 & \cellcolor{worst}53.3 & \cellcolor{worst}46.7 & 60.0 & \cellcolor{worst}61.7 & \cellcolor{worst}38.3 & 60.0 & \cellcolor{worst}43.3 & \cellcolor{worst}58.3 & 70.0 & \cellcolor{worst}52.7 & \cellcolor{worst}48.7 & 61.0 \\
CWE-89 & 51.7 & 53.3 & 65.0 & 60.0 & 41.7 & 70.0 & 61.7 & 40.0 & 55.0 & 63.3 & 33.3 & 70.0 & 56.7 & 46.7 & \cellcolor{worst}90.0 & 58.7 & 43.0 & 70.0 \\
CWE-94 & 58.3 & 46.7 & 70.0 & 70.0 & 31.7 & 70.0 & 70.0 & 35.0 & \cellcolor{highlight}45.0 & 65.0 & 33.3 & 55.0 & 60.0 & 41.7 & 70.0 & 64.7 & 37.7 & 62.0 \\
CWE-200 & 56.7 & 46.7 & \cellcolor{worst}80.0 & 66.7 & 38.3 & 75.0 & 65.0 & 36.7 & \cellcolor{worst}80.0 & 66.7 & 31.7 & \cellcolor{worst}80.0 & 60.0 & 43.3 & 70.0 & 63.0 & 39.3 & 78.0 \\
CWE-416 & 61.7 & 26.7 & 63.3 & 66.7 & 18.3 & \cellcolor{highlight}46.7 & 68.3 & 20.0 & 53.3 & 73.3 & \cellcolor{highlight}13.3 & 53.3 & 70.0 & 26.7 & \cellcolor{highlight}55.0 & 68.0 & 21.0 & \cellcolor{highlight}54.3 \\
CWE-787\&125 & \cellcolor{highlight}71.7 & \cellcolor{highlight}16.7 & \cellcolor{highlight}61.7 & 71.7 & \cellcolor{highlight}15.0 & 50.0 & 73.3 & \cellcolor{highlight}15.0 & 55.0 & 75.0 & 13.3 & \cellcolor{highlight}43.3 & \cellcolor{highlight}73.3 & \cellcolor{highlight}16.7 & 63.3 & \cellcolor{highlight}73.0 & \cellcolor{highlight}15.3 & 54.7 \\
\addlinespace
\midrule
\rowcolor{gray!10}
Avg & 57.1 & 45.0 & 68.6 & 65.4 & 34.8 & 62.8 & 66.3 & 34.2 & 61.4 & 68.0 & 28.6 & 62.2 & 59.5 & 41.9 & 71.9 & 63.3 & 36.9 & 65.4 \\
\bottomrule
\end{tabular}
\footnotesize
\begin{itemize}[leftmargin=*,noitemsep]
\item[$\ast$] Table cells with \colorbox{worst}{red} background denote the worst performance in terms of metrics.
\end{itemize}
\end{table*}
\textbf{Comparison among CWEs.}
We evaluate the performance of multiple LLMs across nine CWEs, covering a broad spectrum of security vulnerabilities. The selected CWEs include \textbf{Input Validation Flaws} (CWE-20, CWE-22), \textbf{Web Security Risks} (CWE-79, CWE-89), \textbf{Memory Safety Issues} (CWE-416, CWE-787\&125), \textbf{Logic \& Configuration Errors} (CWE-94, CWE-200). Due to the identical security measures required for CWE-787 and CWE-125, these vulnerability types are assessed as a combined category.

As shown in Table~\ref{tab:cwe_performance}, Memory Safety vulnerabilities demonstrate the strongest security outcomes, with Qwen2.5Max achieving exceptional results (Avg VIR=13.3\%). Web vulnerabilities exhibit the worst performance, with CWE-79 particularly showing critical limitations with the lowest FCR (48.3--61.7\%) and highest VIR (38.3--58.3\%), indicating fundamental failures in contextual sanitization logic preservation during framework transitions. 

Notably, CWE-20 presents a paradoxical profile, achieving the highest VPR (81.0\%) alongside the lowest non-memory VIR (31.7\%), suggesting models reliably replicate explicit validation patterns without comprehending their security adequacy. CWE-22 shows moderate performance (VIR=39.7\%, FCR=62.0\%). Configuration/logic errors display divergent behaviors --- CWE-94 demonstrates relatively controlled VPR (62.0\%) through standardized mitigation patterns, while CWE-200 exhibits excessive vulnerability preservation (VPR=78.0\%) due to poor detection of implicit resource management assumptions.

The technical profile highlights three tiers of security preservation efficacy: 1) Memory safety vulnerabilities benefit most from architectural language features, 2) Input validation/configuration errors show partial mitigation through syntactic pattern matching, and 3) Web security flaws remain critically problematic due to contextual dependency challenges.
\begin{findingsbox}
\noindent\textbf{Finding 2:} CWEs with explicit syntactic patterns (e.g., input validation) are most preserved in translation (VPR=81.0\%), while context-dependent ones (e.g., XSS) suffer severe security degradation (VIR=48.7\%), showing LLMs prioritize code structure over security semantics.
\end{findingsbox}
\begin{figure*}[t]
  \centering
  \begin{minipage}[t]{0.5\textwidth}
    \centering
    \includegraphics[width=\linewidth, height=5.5cm]{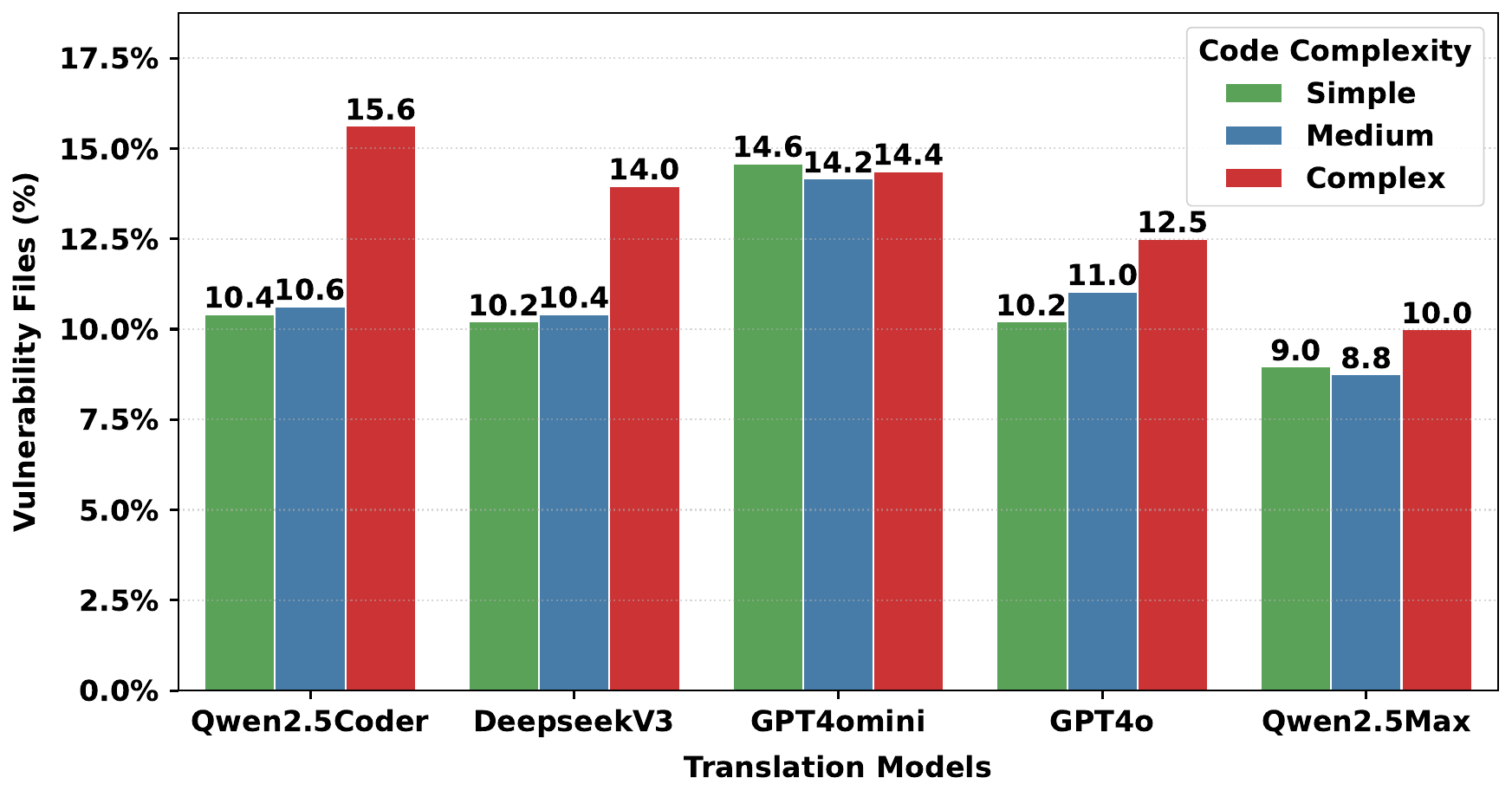} 
    \caption{Complexity Distribution of Vulnerable Translations}
    \label{fig:pillar_vul_dist}
  \end{minipage}\hfill
  \begin{minipage}[t]{0.5\textwidth}
    \centering
    \includegraphics[width=\linewidth, height=5.5cm]{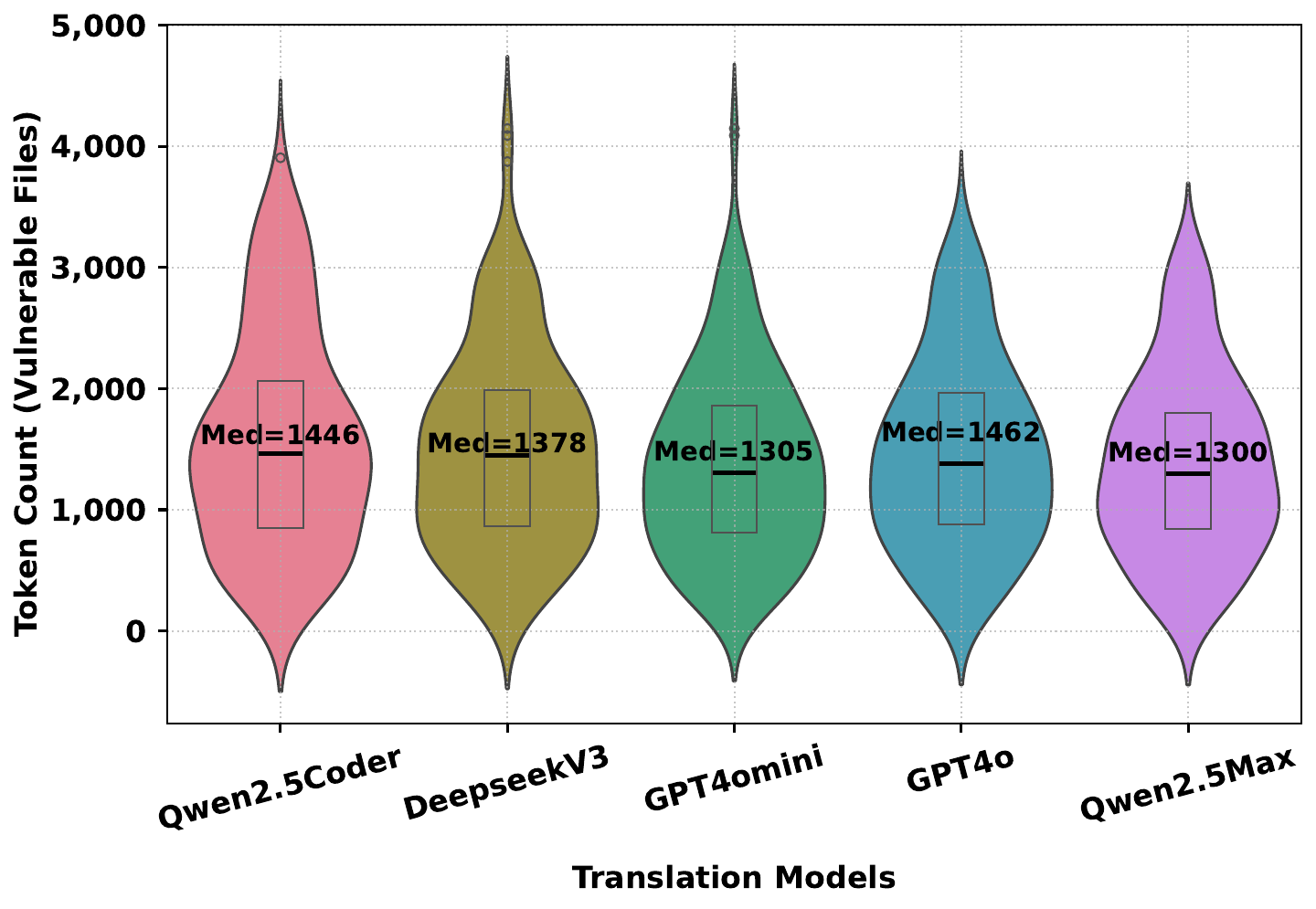} 
    \caption{Token Count Distribution of Vulnerable Translations}
    \label{fig:violin_vul_dist}
  \end{minipage}
\end{figure*}
\textbf{Comparison Among Code Complexity.}
To evaluate the impact of code complexity, we first classify code complexity into three tiers based on token counts: simple (0-950), medium (950-1600), and complex ($>$1600), determined by the 33rd and 66th percentiles of our code corpus. Fig.~\ref{fig:pillar_vul_dist} presents the complexity distribution across LLMs. All models exhibit a positive correlation between error frequency and code complexity, with GPT4omini demonstrating the most uniform distribution while producing the highest error counts -- consistent with its overall bad performance as evidenced by Table~\ref{tab:overall_performance}. DeepseekV3 and Qwen2.5Coder display pronounced sensitivity to code complexity. Notably, Qwen2.5Coder accumulates 15.6\% of its errors in the high-complexity category, suggesting limitations in processing complex code structures. In contrast, GPT4o and Qwen2.5Max maintain relatively stable error distributions across complexity tiers, with Qwen2.5Max achieving the lowest error rate in all categories. This aligns with its best-in-class VIR performance (28.6\% in Table~\ref{tab:overall_performance}), empirically validating its superior capability to preserve security properties.

In Fig.~\ref{fig:violin_vul_dist}, We compare token count distribution of vulnerable code across models using violin plots~\cite{hintze1998violin}. From the perspective of distribution density, Qwen2.5Coder and DeepseekV3 exhibit similar density profiles, both demonstrating higher probability mass in the high token-count regions, indicating comparable limitations when processing complex samples. The remaining three models show relatively uniform distributions. Notably, Qwen2.5Coder displays the largest median value and widest interquartile range, suggesting its errors occur across a broad spectrum of code lengths with particular concentration in more complex samples. Conversely, Qwen2.5Max presents the smallest median and narrowest box width, with errors predominantly clustered around the 1300-token range, consistent with its balanced performance across various complexity levels. GPT4omini's distribution is notably shifted toward lower token counts, with its interquartile range positioned closest to the lower spectrum. This positioning reflects its overall weaker capability compared to other models.
\begin{findingsbox}
\noindent\textbf{Finding 3:} Code complexity directly impacts translation security, with all models showing increased errors in complex code. Qwen2.5Coder and DeepseekV3 show particular sensitivity to complex code structures (VIR of Complex Files = 14.0\% - 15.6\%), while Qwen2.5Max demonstrates more consistent performance across complexity levels.
\end{findingsbox}

\subsection{RQ2:Vulnerable Translation Patterns}
\definecolor{color1}{HTML}{1B6CA8} 
\definecolor{color2}{HTML}{37BC9B} 
\definecolor{color3}{HTML}{F6BB42} 
\definecolor{color4}{HTML}{967ADC} 
\definecolor{color5}{HTML}{E9573F} 

\begin{table*}[ht]
\centering
\caption{Vulnerable Translation Patterns and Frequencies Across Different CWEs (All values in \%)}
\label{tab:cwe_distribution}
\small
\setlength{\tabcolsep}{3.5pt}
\renewcommand{\arraystretch}{1.0}
\begin{tabular}{@{}lccccccccc@{}}
\toprule
\textbf{Vulnerability Category} & CWE-20 & CWE-22 & CWE-79 & CWE-89 & CWE-94 & CWE-200 & CWE-416 & CWE-787\&125 & Total \\
\midrule

\rowcolor{color1!20}
1 Input Validation \& Filtering & \bfseries51.8 & \bfseries62.1 & 27.2 & 33.6 & 35.4 & 21.6 & 9.5 & 28.2 & \bfseries34.9 \\
\rowcolor{color1!10}
1.1 Missing validation logic & 25.4 & 25.3 & 18.9 & 13.2 & 21.1 & 10.7 & 5.4 & 14.1 & 17.3 \\
\rowcolor{color1!10}
1.2 Missing filtering functions & 7.0 & 24.1 & 7.2 & 15.5 & 5.4 & 4.3 & 1.1 & 0.0 & 9.2 \\
\rowcolor{color1!10}
1.3 Validation boundary mismatch & 17.5 & 7.6 & 2.8 & 0.6 & 8.8 & 7.1 & 3.3 & 13.0 & 6.9 \\
\rowcolor{color1!10}
1.4 Normalization mismatch & 0.9 & 5.7 & 0.0 & 0.0 & 0.0 & 0.0 & 0.0 & 0.0 & 0.9 \\

\rowcolor{color2!20}
2 Output Encoding \& Data Protection & 8.0 & 2.6 & \bfseries46.8 & 6.8 & 16.0 & \bfseries40.3 & 0.0 & 1.2 & 17.8 \\
\rowcolor{color2!10}
2.1 Missing encoding layers & 7.0 & 2.5 & 35.0 & 1.1 & 11.6 & 5.7 & 0.0 & 0.0 & 9.3 \\
\rowcolor{color2!10}
2.2 Escaping rule differences & 0.0 & 0.0 & 10.0 & 2.3 & 2.0 & 0.7 & 0.0 & 0.0 & 2.4 \\
\rowcolor{color2!10}
2.3 Inconsistent exception handling & 0.9 & 0.6 & 0.0 & 1.7 & 0.0 & 2.9 & 0.0 & 0.0 & 0.8 \\
\rowcolor{color2!10}
2.4 Sensitive data exposure & 0.0 & 0.0 & 1.1 & 0.6 & 2.0 & 30.7 & 0.0 & 1.1 & 4.6 \\

\rowcolor{color3!20}
3 Security API \& Library Usage & 31.3 & 34.6 & 22.5 & \bfseries58.2 & \bfseries41.0 & 27.3 & 13.1 & 22.4 & 32.7 \\
\rowcolor{color3!10}
3.1 Missing secure API replacement & 19.3 & 8.9 & 11.1 & 43.7 & 17.7 & 16.4 & 2.2 & 0.0 & 16.7 \\
\rowcolor{color3!10}
3.2 API mapping mismatch & 10.5 & 13.9 & 7.8 & 3.4 & 8.8 & 2.1 & 3.3 & 5.4 & 7.1 \\
\rowcolor{color3!10}
3.3 Default behavior differences & 1.8 & 7.6 & 0.0 & 0.0 & 0.0 & 2.1 & 0.0 & 0.0 & 1.5 \\
\rowcolor{color3!10}
3.4 Unsafe function misuse & 0.9 & 3.2 & 2.8 & 16.7 & 15.0 & 6.4 & 6.5 & 15.2 & 8.3 \\

\rowcolor{color4!20}
4 Memory \& Resource Management & 2.7 & 0.0 & 0.0 & 0.7 & 0.0 & 2.2 & \bfseries77.4 & \bfseries48.2 & 10.9 \\
\rowcolor{color4!10}
4.1 Pointer/reference errors & 0.0 & 0.0 & 0.0 & 0.0 & 0.0 & 0.0 & 39.1 & 15.2 & 4.6 \\
\rowcolor{color4!10}
4.2 Bounds operation mismatch & 0.0 & 0.0 & 0.0 & 0.0 & 0.0 & 0.0 & 5.4 & 28.3 & 2.8 \\
\rowcolor{color4!10}
4.3 Lifecycle management failure & 0.9 & 0.0 & 0.0 & 0.6 & 0.0 & 2.1 & 15.2 & 5.4 & 2.2 \\
\rowcolor{color4!10}
4.4 Thread/async risks & 0.9 & 0.0 & 0.0 & 0.0 & 0.0 & 0.0 & 10.9 & 0.0 & 1.0 \\
\rowcolor{color4!10}
4.5 Memory model differences & 0.9 & 0.0 & 0.0 & 0.0 & 0.0 & 0.0 & 7.6 & 2.2 & 0.9 \\

\rowcolor{color5!20}
5 Context \& Framework Behavior & 6.3 & 0.7 & 3.5 & 0.7 & 7.6 & 8.6 & 0.0 & 0.0 & 3.7 \\
\rowcolor{color5!10}
5.1 Missing framework safeguards & 5.3 & 0.6 & 2.2 & 0.6 & 4.1 & 7.1 & 0.0 & 0.0 & 2.6 \\
\rowcolor{color5!10}
5.2 Serialization flaws & 0.0 & 0.0 & 0.0 & 0.0 & 3.4 & 1.4 & 0.0 & 0.0 & 0.6 \\
\rowcolor{color5!10}
5.3 Locale errors & 0.9 & 0.0 & 1.1 & 0.0 & 0.0 & 0.0 & 0.0 & 0.0 & 0.3 \\

\bottomrule
\end{tabular}
\end{table*}
\textbf{Vulnerable Translation Types.}
Prior results demonstrate that even SOTA LLMs exhibit significant limitations in maintaining code security during translation, achieving an average VIR of 36.9\%. Through systematic analysis of vulnerable translations, we aim to identify common security weaknesses in code translation, where all subsequent references to ``errors'' specifically denote security flaws. Employing thematic analysis methodology\cite{cruzes2011recommended}, we categorize 1,549 error cases derived from security-related code translations. We preliminary examine 20\% randomly sampled errors from each CWE to identify primary patterns. Following we develop a refined classification framework with explicit evaluation criteria and representative examples. Security researchers assess each case based on explicitly defined classification criteria, effectively mitigating subjective bias. Major error types include: 
\begin{itemize}[leftmargin=*]
    \item \textbf{Input Validation \& Filtering}: Original input validation or filtering rules are not correctly mapped. Further divided into Missing validation logic, Validation boundary mismatch, Missing filtering functions, Normalization mismatch.
    \item \textbf{Output Encoding \& Data Protection}: Essential encoding layers or data protection mechanisms are improperly implemented across languages. Further divided into Missing encoding layers, Escaping rule differences, Inconsistent exception handling, Sensitive data exposure.
    \item \textbf{Security API \& Library Usage}: Secure API equivalents are either omitted or inaccurately substituted in the target PL. Further divided into Missing secure API replacement, API mapping mismatch, Default behavior differences, Unsafe function misuse.
    \item \textbf{Memory \& Resource Management}: Memory models and resource handling paradigms are incorrectly translated. Further divided into Memory model differences, Bounds operation mismatch, Lifecycle management failure, Pointer/reference errors, Thread/async risks.
    \item \textbf{Context \& Framework Behavior}: Framework-specific security features and contextual protections are lost during translation. Further divided into Missing framework safeguards, Serialization flaws, Locale errors.
\end{itemize}

\begin{figure}[t]
    \centering
    \includegraphics[width=0.5\textwidth, height = 6cm, keepaspectratio]{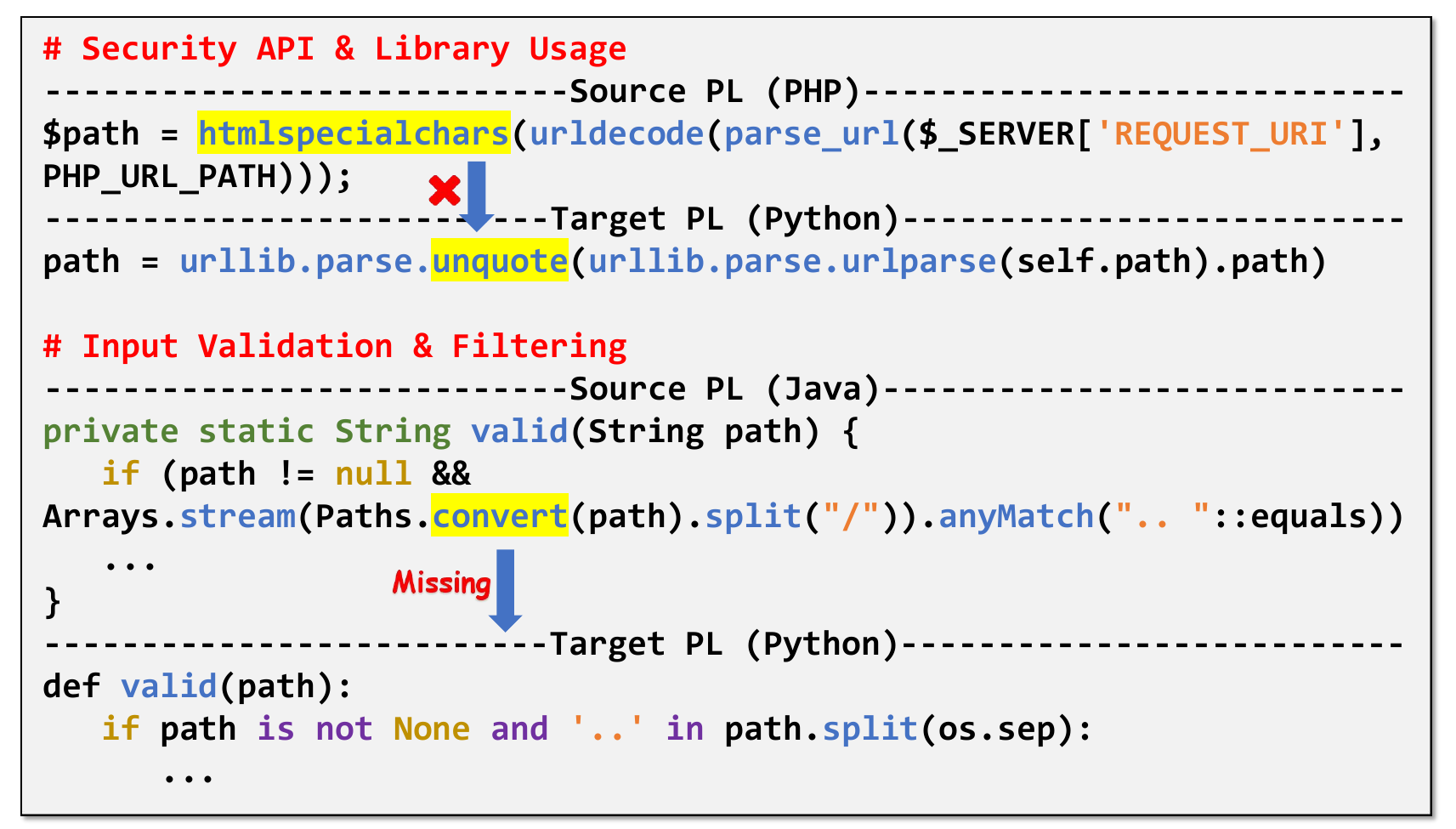}
    \caption{Vulnerable Translation Examples}
    \label{fig:vulnerable_example}
\end{figure}

Fig.~\ref{fig:vulnerable_example} presents concrete cases of vulnerability-introducing translation errors, including: (a) security API mapping mismatch during PHP-to-Python migration, and (b) input validation omission in Java-to-Python translation.

\textbf{Empirical Type Distribution.}
As evidenced by the Table~\ref{tab:cwe_distribution}, vulnerability manifestation shows strong dependence on both the specific weakness type and the underlying language paradigms involved in the translation process.

Memory safety vulnerabilities demonstrate the highest concentration of errors, accounting for 77.4\% and 48.2\% of respective translation failures. This stems from architectural mismatches between source and target memory management models -- particularly in C/C++ to Rust translations where memory operations frequently fail to properly map to ownership semantics. Pointer/reference errors (39.1\%) and bounds operation mismatches (28.3\%) emerge as the most prevalent failure modes, highlighting the challenges in translating low-level memory access patterns to safe constructs.

Web-related vulnerabilities present a distinct profile, with output encoding deficiencies (46.8\%) and API misuse (58.2\%) constituting the primary failure mechanisms. The high prevalence of missing encoding layers (35.0\%) in XSS cases and missing secure API replacement (43.7\%) in SQL injection cases highlights the limitation of LLMs to correctly map security-critical APIs during translation. These weaknesses prove particularly severe when translating between web ecosystems with divergent protection mechanisms.

Input validation flaws exhibit more distributed error patterns, though still dominated by validation logic omissions (25.3-25.4\%) and filtering function omissions (24.1\% in CWE-22). The relative uniformity of these errors across PL pairs suggests that while validation constructs are syntactically easier to translate, their semantic adequacy frequently degrades during translation -- particularly for boundary condition checks (17.5\% in CWE-20) and normalization requirements (5.7\% in CWE-22). Configuration and logic errors reveal an inverse pattern, with errors primarily emerging from improper handling of implicit behaviors rather than explicit code constructs. Sensitive data exposure (30.7\% in CWE-200) and Secure API omission (17.7\% in CWE-94) dominate these categories, reflecting LLMs' difficulty in preserving environmental security assumptions and default-safe configurations.

\begin{figure}[t]
    \centering
    \includegraphics[width=0.5\textwidth, keepaspectratio]{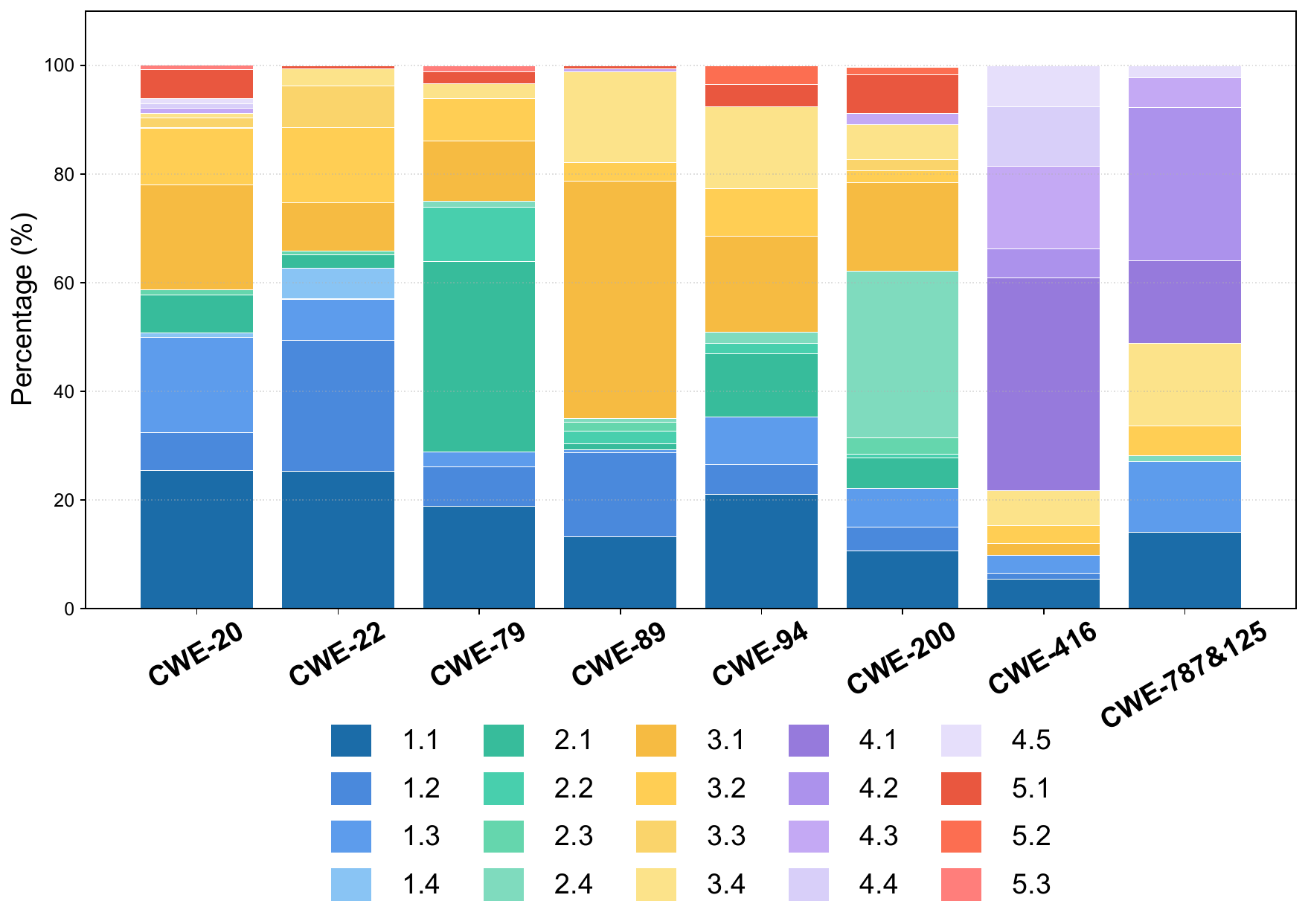}
    \caption{Vulnerable Type Distribution Among CWEs}
    \label{fig:cwe_dist}
\end{figure}

The stacked percentage bar chart (Fig.~\ref{fig:cwe_dist}) reveals critical vulnerable patterns across different CWEs. While input validation flaws (34.9\%) and API misuse (32.7\%) dominate the overall error distribution, the distribution highlights three fundamental translation weaknesses: First, the persistence of input validation errors reveals models' tendency to preserve code structure while losing security semantics. Second, the prevalence of API misuse (32.7\%) underscores the difficulty in mapping equivalent secure functions across PLs. Third, the significant rate of output encoding failures (17.8\%), especially in web contexts (46.8\% for CWE-79), demonstrates models' limited understanding of framework-specific security requirements.
\begin{findingsbox}
\noindent\textbf{Finding 4:} Input validation (34.9\%) and Security API (32.7\%) constitute the two most prevalent sources of translation-induced vulnerabilities. This concentration reveals that LLMs prioritize syntactic correctness over security semantics when translating critical code constructs.
\end{findingsbox}
\subsection{RQ3:Vulnerable Translation Causes and Risks}
\textbf{Vulnerable Translation Causes.} 
The security attributes of programming languages fundamentally influence the security of code translation. A case study on C to Rust code migration reveals the interplay between language security features and implementation strategies. As Rust's ownership model theoretically eliminates memory safety flaws, 16.7\% of translated cases introduced vulnerabilities (shown in Table~\ref{tab:cwe_performance}) , while 58.3\% -- as measured in this study -- failed to apply Rust's safety mechanisms. Data highlights two patterns: translations adhering to Rust's safeguards achieved superior outcomes (0\% VIR for safe-only code), whereas those bypassing safety checks worsened results (16.7\% VIR), proving security outcomes depend on framework adaptation rather than functional correctness. The study disproves the assumption of automatic security inheritance through language translation. Ultimately, Target language's security properties define theoretical security boundaries, but their practical effectiveness depends on rigorous adherence to safeguard mechanisms during translation.

The prevalence of input validation and security API vulnerabilities in translated code exposes fundamental limitations in how LLMs reconcile syntactic and functional correctness.

Input validation flaws demonstrate LLMs' propensity for syntactic preservation over semantic validation. While models successfully translate explicit validation constructs with structural equivalence, most of these translated safeguards exhibit semantic decay in boundary enforcement. This explains the high vulnerability preservation rates (VPR=81.0\%) for CWE-20. This ``validation equivalence fallacy" arises because LLMs optimize for structural fidelity without modeling the security implications of language-specific type systems.

Security API misuse reveals a parallel failure mode in semantic mapping. When translating Java's PreparedStatement to Python DB APIs, some cases incorrectly substitute parameterized queries with string concatenation -- a syntactically valid but semantically dangerous pattern. The models recognize API call structures but fail to comprehend their security purpose.
Input Validation and Security API failure patterns stem from LLMs' training paradigm that prioritizes syntactic congruence over security semantics. 

\textbf{Real-world Risks of Vulnerable Translations.}
To assess the real-world impact of vulnerable translations, we conduct a developer study using a questionnaire based on translation examples from our code corpus. The study involves 30 participants categorized into three groups according to their programming and security expertise. Our survey focuses on three core aspects of vulnerability detection in LLM-based code translation:
\begin{itemize}
\item Q1: What types of issues in translated code do you find most challenging to verify and debug?
\item Q2: Vulnerability detection capability assessment through eight translation samples.
\item Q3: Which vulnerability type proved most difficult to identify based on your evaluation experience?
\end{itemize}
\begin{figure}[t]
    \centering
    \includegraphics[width=0.7\textwidth, height=5cm, keepaspectratio]{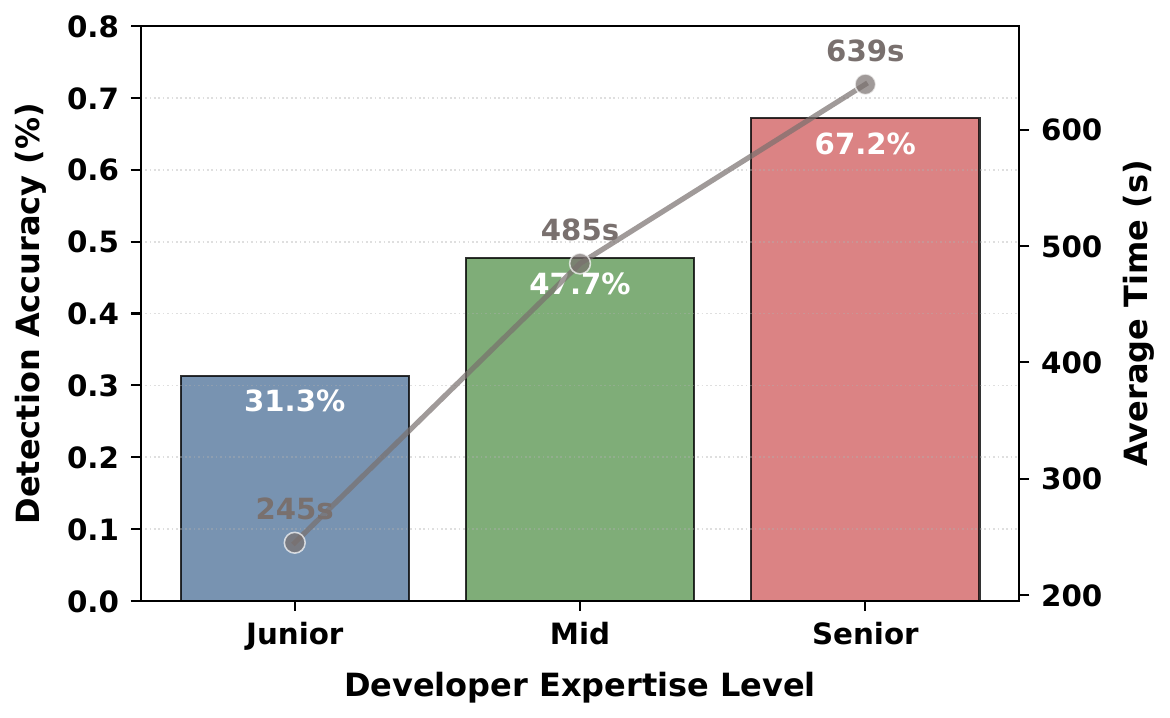}
    \caption{Developer Performance in Translation Vulnerability Detection}
    \label{fig:questionnare}
\end{figure}

Our survey results reveal critical risks posed by vulnerable translations in real-world development scenarios. The data demonstrates that security vulnerabilities in translated code present significant detection challenges, with 90\% of developers identifying them as the most difficult issues to verify -- far surpassing syntactic errors (93.3\% agreement on being easiest). This perception-risk mismatch highlights a dangerous blind spot, where the most severe vulnerabilities are also the hardest to identify. Fig.~\ref{fig:questionnare} exposes concerning accuracy limitations across all expertise levels. While senior developers perform best (67.2\% accuracy), their detection rates remain alarmingly low given the security-critical nature. More troubling is the observed inverse relationship between expertise and evaluation efficiency -- senior developers require 161\% more time (639s vs 245s) than juniors. This suggests that vulnerable translations force experienced developers into defensive scrutiny patterns, while less experienced ones may unknowingly accept flawed translations.

When asked to identify the most challenging vulnerability types, developers equally prioritize two critical risks: security API mismatches (43.3\%) and language paradigm transitions (36.7\%). This near-even split indicates that vulnerable translations threaten systems through both localized API misuse and systemic safety model violations. The prolonged verification times (mean 478s) further compound these risks, as such thorough reviews prove difficult to maintain in practice.

The complete questionnaire is available via the anonymous link provided by our team\cite{STEDanonymouslink}.
\begin{findingsbox}
\noindent\textbf{Finding 5:} Vulnerabilities in LLM-translated code are highly severe and difficult to detect, with an average developer accuracy of only 49.6\%. Senior developers, while more accurate (67.2\%), require 161\% more time than juniors, revealing an efficiency-accuracy tradeoff.
\end{findingsbox}
\subsection{RQ4:Mitigation Effectiveness}
To address security vulnerabilities in LLM-based code translation, we implement and evaluate two mitigation strategies: a naive security-aware approach and a Retrieval-Augmented Generation (RAG) method~\cite{lewis2020retrieval}.

The \textbf{naive security-aware approach} enhances standard translation prompts by integrating explicit security requirements, instructing models to preserve functional correctness while avoiding common vulnerability patterns. While providing a foundational layer of security guidance, its advice remains generic and not specifically tailored to the semantic content of the input code.

\begin{figure}[t]
    \centering
    \includegraphics[width=0.5\textwidth, keepaspectratio]{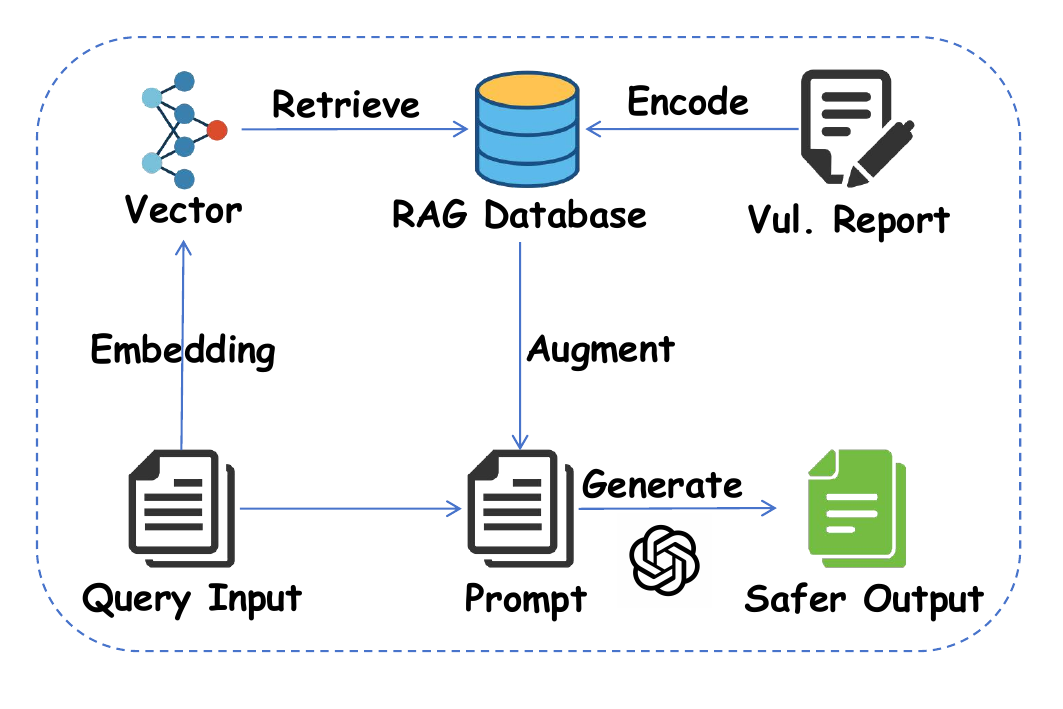}
    \caption{Retrieval-Augmented Generation Method Overview}
    \label{fig:RAG}
\end{figure}

In contrast, our \textbf{RAG-based solution}, as shown in Fig.~\ref{fig:RAG}, employs a more sophisticated pipeline to deliver context-aware security hardening. The system is built upon a meticulously constructed knowledge base of vulnerability analysis reports. This specialized knowledge base is curated from a corpus of code samples that are previously identified through human analysis as containing security vulnerabilities. To ensure comprehensive coverage, we employ a stratified sampling strategy across diverse CWE types. Each selected code sample is accompanied by a manually authored, in-depth analytical report detailing the nature of the vulnerability, its root cause, and potential ramifications. The knowledge base is constructed in accordance with established best practices, utilizing a maximum chunk length of 1,500 words and an overlap of 300 words, as recommended by platforms like Coze. Under this configuration, the textual content is partitioned into 128 distinct chunks, a structure designed to optimize retrieval coverage and precision.

Our RAG framework operates through a meticulously designed two-phase architecture that seamlessly integrates offline preparation with online retrieval and generation. During the offline processing phase, we establish the foundational knowledge base by employing a pre-trained sentence transformer model (all-MiniLM-L6-v2~\cite{wang2020minilm}) to encode all code snippets from our vulnerability repository into dense vector representations. This embedding process is optimized through batch processing and L2 normalization, ensuring both computational efficiency and optimal vector space geometry for subsequent similarity measurements. The resulting embedding matrix is persistently stored alongside the original vulnerability metadata, creating a comprehensive knowledge repository that forms the cornerstone of our retrieval mechanism.

In the online phase, the system initializes by loading the pre-computed embeddings and vulnerability data into memory, ensuring rapid response capabilities. When presented with a new code translation task, the framework first generates an embedding vector for the input code using the same sentence transformer model. This query embedding then undergoes a sophisticated similarity assessment against the entire stored vector database through cosine similarity calculations, efficiently identifying the three most semantically analogous code cases that exceed a predetermined similarity threshold of 0.5.

RAG system can architecturally integrate the retrieved security intelligence directly into the prompt. This meticulously crafted prompt synthesizes the source code, fundamental translation requirements, and a specialized Security Considerations section that enumerates specific vulnerability types, severity assessments, and detailed human-authored analysis reports corresponding to the retrieved cases. This contextual enrichment transforms the prompt from a mere translation directive into a comprehensive security-aware generation framework. Finally, this richly contextualized prompt is presented to the backend LLM, enabling the translation process to be informed by precise vulnerability patterns associated with semantically similar code structures.

\begin{table}[t]
\centering
\caption{Performance Comparison of Different Strategies}
\label{tab:mitigation_improvement}
\begin{tabular}{lcccc}
\toprule
& \multicolumn{2}{c}{\textbf{GPT4omini}} & \multicolumn{2}{c}{\textbf{DeepseekV3}} \\
\cmidrule(lr){2-3} \cmidrule(lr){4-5}
\textbf{Strategy} & \textbf{VIR} & \textbf{FCR} & \textbf{VIR} & \textbf{FCR} \\
\midrule
Baseline       & 100.0\% & 100.0\% & 100.0\% & 100.0\% \\
\cmidrule(r){1-1} \cmidrule(lr){2-3} \cmidrule(lr){4-5}
Naive Prompt   & 74.6\%  & 89.6\% & 88.8\%  & 92.5\% \\
Change    & +25.4\%  & -10.4\% & +11.2\%  & -7.5\% \\
\cmidrule(r){1-1} \cmidrule(lr){2-3} \cmidrule(lr){4-5}
RAG Prompt     & 67.2\%  & 85.1\% & 66.7\%  & 88.1\% \\
Change    & +32.8\%  & -14.9\% & +33.3\%  & -11.9\% \\
\bottomrule
\end{tabular}
\end{table}

Our evaluation on 67 security-centric code samples demonstrates progressive improvement across mitigation strategies, with VIR consistently declining from baseline to RAG approaches (Table~\ref{tab:mitigation_improvement}). The naive prompt reduces VIR by 25.4\% for GPT4omini and 11.2\% for Deepseek compared to baseline, confirming that explicit security constraints in prompts provide moderate protection. However, the RAG strategy delivers significantly better outcomes, achieving an additional 7.4\% VIR reduction over naive prompts for GPT4omini and 22.1\% for Deepseek. Notably, both models exhibit comparable sensitivity to RAG-based mitigation, with GPT4omini showing a 32.8\% improvement and Deepseek a 33.3\% improvement, suggesting that external contextual guidance can effectively compensate for varying model capabilities.

To assess functional correctness, we conducted an additional evaluation on a separate set of 67 samples, as the VIR test samples were inherently ill-suited for this metric. The results reveal a trade-off: while mitigation strategies enhanced security, they incurred a cost to functional correctness. The Naive Prompt led to an FCR reduction of 10.4\% for GPT4omini and 7.5\% for Deepseek. The RAG strategy, despite its superior security performance, resulted in a further decline in FCR to 14.9\% and 11.9\% for the respective models. This inverse relationship highlights the inherent tension between security and functionality. Crucially, the RAG approach demonstrates a more favorable trade-off, achieving the most significant security gains (32.8\%/33.3\% VIR reduction) for a moderate and controlled decrease in functional correctness, as compared to the baseline. This performance convergence implies that well-designed mitigation strategies can bridge inherent security awareness gaps across different LLMs while managing the impact on their functional capabilities.
\begin{findingsbox}
\noindent\textbf{Finding 6:} Mitigation strategies demonstrate hierarchical effectiveness, with RAG prompts achieving substantial improvements (32.8\%–33.3\%) over naive security prompts (11.2\%–25.4\%). This highlights the critical role of contextualized security knowledge integration in enhancing secure code translation.
\end{findingsbox}
\section{Discussion}
\subsection{Threats to Validity}
\textbf{Threats to internal validity} mainly arise from potential data leakage and the use of manual processes in data collection and evaluation~\cite{zhou2024large}. Although some LLMs may have been trained on source code from our collected samples, they lack access to the corresponding correct translations, ruling out task-specific fine-tuning or reinforcement. Our results show that even when exposed to security-aware functions, LLMs still make frequent errors in translating security APIs, suggesting minimal impact from data leakage. To ensure reliability in manual procedures, we follow standardized guidelines and involve multiple annotators during both data preparation and evaluation to reduce bias and human error. For RQ2, we employ multiple experienced researchers under a double-blind review process, enhancing the robustness of our results.

\textbf{Threats to external validity} relate to the generalizability of our findings across programming languages and security contexts~\cite{zhang2024hard, liu2024exploring}. Our study focuses on five languages and nine CWE categories, which naturally limits its scope compared to the wide variety of domain-specific languages and vulnerabilities. However, the selected languages are among the most popular in industry~\cite{TIOBE} and represent diverse programming paradigms relevant to secure coding~\cite{whitesource}. The chosen CWEs are drawn from the MITRE Top 25\cite{cwetop25}, covering common security scenarios. Furthermore, we conducted a thorough analysis of all security-related translation errors using thematic analysis to enhance the robustness and applicability of our results. Thus, we believe the threats to external validity are minimal within the scope of this work.
\subsection{Practical Value}
This paper presents the first comprehensive evaluation of security issues in LLM-based code translation, uncovering security risks introduced during file-level translation. Based on our findings, our work offers the following practical value.

\textbf{Value for Researchers.} 
We provide a comprehensive security-centric evaluation methodology and empirical framework that can facilitate future research in this domain. This study fills a critical gap in security evaluation within code translation research by establishing rigorous evaluation protocols. Additionally, we introduce the first taxonomy of security-related translation errors, offering a foundational reference for future studies. Our vulnerability analysis and the mitigation approach presented in RQ4 provide actionable insights for addressing security challenges in LLM-based code translation.

\textbf{Value for Practitioners.} 
Our experimental results offer practical guidance for developers using LLMs in real-world code translation tasks. The overall performance assessment in RQ1 informs model selection decisions, while the vulnerability patterns identified in RQ2 highlight critical areas requiring manual review. The developer survey in RQ3 emphasizes the importance of maintaining caution about security risks in LLM-based translation and provides concrete evidence of the challenges developers face in detecting translation-induced vulnerabilities.
\section{Conclusion}
This study presents the first comprehensive empirical investigation into the security implications of LLM-based code translation. Through rigorous evaluation across five programming languages, nine CWE types, and five state-of-the-art LLMs, our research reveals significant vulnerability introduction and preservation issues in automated code translation. The manual analysis of translation errors and the development of mitigation strategies provide valuable insights into both the limitations and potential improvements for secure code translation.

Based on our findings, future work may include:
(1) Developing integrated prompting frameworks that combine static analysis with real-time security context retrieval to better identify and handle security-sensitive code during translation.
(2) Exploring security-specialized fine-tuning approaches using high-quality translation pairs annotated with vulnerability patterns, potentially leveraging reinforcement learning from security feedback to align models with secure coding practices.
(3) Constructing standardized security benchmarks for code translation evaluation, incorporating multi-granularity test cases, dynamic analysis validation, and comprehensive metrics to reliably assess the security integrity of translated code.

\bibliographystyle{plain}
\bibliography{ref.bib}
\end{document}